# ON THE INTERPRETATION OF THE FOUNDATIONS OF QUANTUM MECHANICS

## V. E. Egorushkin

Tomsk State University, Institute of Strength Physics and Materials Science of the SB RAS

This study discusses the quantum behavior of a particle, which is controlled by fluctuations in the physical space-time (ST) variables, rather than provides a novel interpretation of quantum theory. The fluctuations, i.e., inhomogeneities in a homogeneous phase ST, are prescribed by their probability. They determine the reciprocal space and correlate with the correlation entropy different from zero. Alongside with the minimum entropy (S) (k is the Boltzmann constant), action (s) (h is the Planck constant), and the presence of the Winn-Ehrenfest adiabatic invariant (AI), the fluctuations require the Gilbert information (probabilistic) space linking the physical and the reciprocal ST. Physical quantities in the information space are represented by linear Hermitian operators, which is due to the entropy production in the presence of an AI. Evolution of a quantum system is described by the wave functions having the meaning of information concerning all virtually possible states of a quantum particle. The wave functions are the solutions to the Schrödinger equation and represent a navigation 'roadmap' for the particle to follow. A quantum system is in fact a classical Hamiltonian system in the space of $C_n, C_n^*$ coefficients of the wave function decomposition with respect to the operator eigenfunctions. These coefficients prescribe the particle trajectorys in the $C_n, C_n^*$ space.

It is the linearity and the Hermitian nature of the operators which determine the trajectory and the superposition principle (not vice versa) in case of the wave behavior of fluctuations. The uncertainty principle, resulting from $\Delta s \geq h$, reflects correlation of the fluctuations and, hence, their nonlocality.

This study discusses the wave function phase, the Berry phase and its relationship to quantization, discriminability of states and macroscopic quantum effects caused by localization of the particle, followed by a possible entropy change during its transition into a new thermodynamic state.

This work addresses interference: it is the information waves which interfere; the particles follow their roadmap, and the measurement of their coordinates introduces an additional uncertainty into the momentum. It is not particles (matter) which manifest these wave properties but fluctuations of the ST coordinates. The duality of fluctuations is as follows: on the one hand (for coherent states) $\Delta x = h/p$, while on the other hand, $\Delta x$ is the trajectory difference of interfering waves, which, due to the condition of interference, is equal to the wavelength – $\lambda$. The condition $\Delta x = \lambda$ ensures interference even for a single quantum particle.

Physical characteristics corresponding to the fluctuating variables – energy, momentum, etc. determine the magnitude of the respective fluctuations rather than the wave properties of matter. Matter possesses no wave properties.

This work also discusses the difference between the objective information and knowledge.

---

**Corresponding author:** Valery Egorushkin, Professor, research fields: nanomaterials (electrical properties, thermal properties), plastic strain, disorder systems. E-mail: val110@mail.ru.

# 1. Introduction.

Despite a hundred years of the history of quantum mechanics (QM), an interest in its foundations is ever increasing. The annular symposia and great progress in the experimental investigation of isolated quantum objects, numerous technological applications, such as quantum computing, cryptography, etc., continuously stir up this interest [1, 2].

It is believed that the problem lies behind quantum measuring as described by the numerous available studies. Unfortunately, their interpretations of QM ranging from the Copenhagen to the informational interpretations of [3-17] fail to properly present the foundations of QM: the principles of uncertainty, superposition, complementarity, corpuscular-wave dualism, sense of the wave-function, Schrödinger equation, wave function collapse during measurement, and quantum interference, and make them clear even to undergraduate students of the physics departments.

According to most scholars, the problem stems from the fact that QM lacks the basic principles that could form certain references for the quantum theory in the same way as do, e.g., the constancy of velocity of light (c) and the invariance of interval in the special relativity theory (SRT). On the other hand, there is a constant in the quantum theory which is in a way similar to the velocity of light in SRT; it is Planck's constant – h, denoting the minimum change of action – $\Delta s$, hence we take that

$$\Delta s \geq h.$$

Moreover, there is an adiabatic invariant $\nu/T$ determined by Ehrenfest [18] and Winn [19], which is expressed via the ratio of two constants: Boltzmann's – k and Planck's – h.

The investigations into the features peculiar to the quantum world, such as the Einstein-Podolsky-Rozen paradox (EPR), entangled states, Bell's inequality, decoherence, possibility of a classical outcome of quantum events, nonlocality, etc., have made such a long step away [20], that seldom come back to the origin of quantum behavior – the existence of h and adiabatic invariant.

Being a university professor of the physics department for a large number of years, lecturing on atomic and nuclear physics, quantum mechanics, physics of non-equilibrium phenomena, [21], and being familiar with the current studies in these fields, every year I have to explain to the undergraduates the origin of quantum description and its basic principles. While on the mathematical side a logical reasoning would yield correct results, on the physical side the interpretations appear to be nearly dogmatic.

It is 'the lack of real comprehension' which motivated my attempt to analyze quantum peculiarities, relying on the two abovementioned factors  – 'principles' and the resulting conclusions pertaining to the wave behavior of fluctuations, in particular, the principles of uncertainty, superposition, complementarity, sense of the wave-functions and their phases, dualism, Schrödinger equation, interference, and macroscopic quantum behavior.

## 2. Least action and uncertainty principles.

### 1) Uncertainty relation

Let us use the fact that Planck's constant h is the minimum mechanical action, i.e., no variation of an action (s) can be lower than h (principle of least action)

$$\Delta s \geq h. \qquad (*)$$

Writing the term for mechanical action via conjugate variables (p,x) in a one-dimensional case, from (*) we arrive at

$$\Delta(p \cdot x) = x\Delta p + p\Delta x \geq h \qquad (**)$$

Assuming $\Delta p, \Delta x \neq 0$, we obtain

$$\Delta p \Delta x \geq \left(\frac{W_x W_p}{W_x + W_p}\right) h, \qquad (1)$$

where $W_x = \frac{\Delta x}{x}$ is the probability that coordinate x would fluctuate (have an uncertainty) - $\Delta x$, and $W_p = \frac{\Delta p}{p}$ is the probability of uncertainty $\Delta p$ of momentum p.

Quantity $W = \frac{W_x W_p}{W_x + W_p}$ is determined by the correlation of fluctuations $\Delta x$-$\Delta p$ and represents a reduced probability, which is equal to the ratio of probability $W_x W_p$ of superposition ($\Delta x \cap \Delta p$) of independent, $\Delta x$, $\Delta p$ to the probability ($W_x + W_p$) of combination ($\Delta x \cup \Delta p$) of inconsistent $\Delta x$, $\Delta p$.

Since $\frac{W_x + W_p}{W_x W_p} \geq 2$, then if (1) is fulfilled when being equal to two, then it will be satisfied for all $\frac{W_x + W_p}{W_x W_p} > 2$, i.e., (1) can be rewritten

$$\Delta p \Delta x \geq \frac{h}{2}, \qquad (2)$$

as Heisenberg's uncertainty relation.

### 2) Einstein formula for fluctuations. Binding entropy.

Let us return to (1). Take ln on both sides of (1) and multiply it by the Boltzmann constant. Then for the equality (of coherent states) we have

$$k \ln \frac{\Delta x \Delta p}{2\pi h} = k \ln W \qquad (3)$$

By definition, on the left-hand side of (3) would be the entropy (ΔS) due to fluctuations Δx, Δp, i.e.,

$$\Delta S = k \ln W \qquad (4)$$

is the Einstein formula for fluctuations (in the general case ΔS is no less than $k \ln W$).

Since
$$\Delta S = k \left( \ln W_x W_p - \ln(W_x + W_p) \right), \qquad (5)$$

then, similarly to the binding energy, <u>ΔS – the binding entropy -</u> is due to inconsistency of Δx, Δp, i.e., the correlation between them. It is impossible to avoid correlations within the framework of a quantum description: ΔS≠0 under no circumstances without an external force. This dictates nonlocality of quantum behavior – wave behavior of fluctuations, having so far nothing to do with the entanglement of states and Bell's inequalities. It should be underlined here that it is the binding entropy that determines the derivation of (1) and (2).

Relations similar to (1) and (2) will also occur for $\Delta E \Delta t \geq \left( \frac{W_E W_t}{W_E + W_t} \right) h$ and other conjugate variables.

### 3) Wave nature of fluctuations.

The 'miracle' of appearing h (phase space discontinuity) gives rise to fluctuations of physical quantities and space-time coordinates. These fluctuations are not associated with either the measurement or the observer (provided that the appearance of h itself is not due to these circumstances).

It is clear from relations (1) – (3) that h determines the relationship of fluctuations of physical quantities and space-time images with <u>information.</u>

The development of fluctuations 'deteriorates' the homogeneity of space (Δp) and time (ΔE), and the isotropy of space (ΔM). The requirement of maximum homogeneity of ST determines Δp→0 and ΔE→0; and proceeding from relation (2) it is erroneous to argue that Δx and Δt are undefined.

In order to understand what happens to the space-time fluctuations, one should resort to relation (\*\*) rather than (2), from which for Δp→0, we have
$$\Delta x \geq h/p \qquad (+)$$

In the case of an equality, (+) turns into a definition of the de Broglie wavelength

$\Delta x = x = h/p$ and $p = \hbar k$ ( $k = \frac{2\pi}{\lambda}$ ). Similarly, $\Delta t = h/\Delta E$ and $E = \hbar \omega$ ( $\omega = \frac{1}{2\pi \Delta t}$ ).

Time-space fluctuations correspond to the length and frequency of a certain wave. Note that coordinates x and time points t, i.e., the trajectory, remain unknown. In a quantum system, motion is associated with k, ω rather than with x, t, hence, the uncertainty principle is a necessary condition for complimentarity to appear: this indicates the wave character of variations in the spatial characteristics without specifying the physical sense of these waves. Furthermore, it is well known that minimizing the uncertainties of the energy density of a single-electron atom with respect to $\Delta x$, we obtain the following estimates: $\Delta x \sim \frac{\hbar^2}{me^2} \sim 0.5 \dot{A}$, $\Delta E \sim \frac{me^4}{\hbar^2} \sim 13.6$ eV, the latter being a possible spacing between the electron and the nucleus and the ionization energy in the hydrogen atom, respectively; if $\Delta x = c\Delta t$, then $\Delta E = \Delta mc^2$.

Thus, the ratio of uncertainties reflects the relationship of 'space-time images', the laws of conservation of motion, and the process of motion of a quantum particle irrespective of the variation in the characteristics of this motion. This is entirely in the spirit of Einstein stating that there is something like a real state of a physical system which exists objectively, irrespective of any measurement or observer.

From my standpoint, relating h to the measurement and the observer is the same as relating to them Boltzmann's constant k (a quantum of entropy) or Avogadro's number, which possess no less meaning than h does. Thus, the question of why h appears is still open.

For the sake of being objective, let us note that expression (+) relates fluctuations to momentum and energy only, without indicating their wave nature. We merely used 'resemblance' with the De Broglie formula. When addressing interference, in what follows $\Delta x$ will be unambiguously related to the length of interfering waves.

## II. Schrödinger equation and wavefunctions.

### 1. Hilbert information space.

In a homogenous physical ST, both $\Delta x \neq 0$ and $\Delta t \neq 0$ can equally exist in any region, and a relation such as (4) is therefore satisfied for densities S(x,t), W(x,t). The emergence of fluctuations (and their wave nature) is associated with the entropy production $\frac{dS(x,t)}{dt}$ and variation $\frac{dS(x,t)}{dx}$.

Let us analyze the entropy variation taking into account the existence of h and Winn-Ehrenfest adiabatic invariant. For the densities from (4) we have

$$S(x,t) = k \ln W(x,t) \qquad (5)$$

Let us derive the entropy production given that there is a certain region of action inhomogeneity ($\Delta x \cdot \Delta p$), which determines $\frac{dS}{dt}$ so that $\frac{dS}{dt} = \frac{dS}{ds} \cdot \frac{ds}{dt}$, as it is generally accepted in locally equilibrium processes.

Let us use the fact that

$$\frac{dS}{ds} = \lim_{\Delta S \to \hbar} \frac{\Delta S}{\Delta s} = \frac{k}{\hbar} = \frac{\omega}{T} = \frac{E\omega}{ET} = \frac{S}{s} \tag{6}$$

is the adiabatic invariant.

Therefore, the emergence of h in the system's quantum behavior is due to the existence of an entropy quantum – k.

Note that the classical principle of least action $\delta s = 0$ is applied only if $\Delta S = 0$, i.e., when h and k can be neglected in favor of larger quantities such as s, S.

Considering (6), we immediately obtain a relation for action and probability

$$s(x,t) = \hbar \ln W(x,t). \tag{7}$$

Since $\frac{ds}{dt} = -H$ are the Hamiltonian functions, then from (7) we arrive at

$$-H(x,t) = \hbar \frac{d}{dt} \ln W(x,t), \tag{8}$$

i.e., the system's energy is spent on production of <u>information</u> about it, while ℏ acts as a transfer coefficient between the physical quantity – energy – and the information on space-time fluctuations, which have wave nature (λ,ω) and characterize motion of a quantum particle. The quantities responsible for variations in Δx,Δp, Δt, ΔE,… can appear as oscillations or rotations (and their respective Fourier images). What are these quantities related to probability W(x,t)?

In a physical (phase), Euclidean, homogenous space, where Δx and Δt are mere characteristics of the waves, and Δp, ΔE,… are equal to zero, both these oscillating quantities related to W(x,t) and W(x,t) as such are out of the question.

Inherently $W(x,t) = \frac{W_x}{\left(W_x + W_p\right)^{1/2}} \cdot \frac{W_p}{\left(W_x + W_p\right)^{1/2}}$ represents a certain 'bilinear form' that in a physical space is a null equation, since $W_p \sim \Delta p \approx 0$.

In order to analyze the right-hand side of (8), we need to consider a different space – quantum information space – for which there are defined (propagating) waves representing the probability density W(x,t). Their wave characteristics are fluctuations Δx,Δt, and the space proper would be 'isotopic' with respect to the physical phase space.

This information space would be the <u>space of functions</u> representing waves. This space would also be space dual [22]. If we introduce quantum (Matsubara) time $t \to it$, then the information quantum ST with a normal coordinate system and unit transformations would be represented by the complex Hilbert space [21].

## 2. Operators of physical quantities in the information ST. Quantum-mechanics description of virtual states.

Let us represent W(x,t) in the information space as $W(x,t) = \psi(x,t)\psi^*(x,t)$. In (8), discriminating between $\psi$ and $\psi^*$, and using for S using complex variables including the quantum time $it$, we obtain

$$H\psi(x,t) = i\hbar \frac{\partial \psi(x,t)}{\partial t}; \quad H\psi^*(x,t) = -i\hbar \frac{\partial \psi^*(x,t)}{\partial t}. \qquad (9)$$

From (9) follows that <u>energy</u> in the case of quantum behavior can be represented by a linear operator $i\hbar \frac{\partial}{\partial t}$ (Hermitian form) in the Hilbert space – a generalization of the Euclidean space of finite vectors with respect to the functions with a finite norm for $\int |\psi|^2 dx = 1$ (or infinite with respect to a different normalization condition). It is the transfer to the quantum time which determines the Hermitian character of operators and, as will be clear further, the classical Hamiltonian essence of quantum dynamics.

Introducing instead of $\frac{d}{dt}$ a substantial derivative $\frac{\partial}{\partial t} + \upsilon \frac{\partial}{\partial x}$, in the same way as before, equating the left- and right-hand sides of the expressions for $\frac{\partial}{\partial t}$ and $\frac{\partial}{\partial x}$ respectively, along with (9) we obtain

$$p\psi(x,t) = -i\hbar \frac{\partial \psi(x,t)}{\partial x}. \qquad (10)$$

Momentum in the space $\psi(x,t)$ has a corresponding linear Hermitian operator ($-i\hbar \frac{\partial}{\partial x}$) and, therefore, momentum is determined (determines) by a spatial change $\psi(x,t)$ – information inhomogeneity.

Thus, <u>quantum dynamics is determined by the space-time inhomogeneity of information</u>.

Let us underline that (9) is not an equation. It only determines the energy operators in the Hilbert space. However, if we represent H as an operator $\hat{H}$, expressing it via operators $\hat{p}, \hat{x},...$, using the <u>correspondence principle</u>, then (9) will turn into a Schrödinger equation.

In a quantum case, instead of a phase ST with a moving particle, there is a different – 'isotopic' information ST– due to fluctuations of the space-time motion coordinates that result from the fact that no change of action can be smaller than h, and no change of entropy can be smaller than k. It should be noted that similarly to the mechanical state of a particle in the motion space it is determined by the integral of motion – conserved physical quantities E, p, N,…, and the information space ensures that the mechanical state is determined in the same fashion – via the integrals of motion – mean values of physical quantities E, p, M,… as follows:

$$p = -i\hbar \left\langle \psi \left| \frac{\partial}{\partial x} \right| \psi \right\rangle; \quad E = i\hbar \left\langle \psi \left| \frac{\partial}{\partial t} \right| \psi \right\rangle \quad \text{(in Dirac's notations)}. \tag{11}$$

Since the operators of physical quantities are linear and Hermitian, then an arbitrary function $\psi(x,t)$ can be represented as a <u>superposition</u> of eigenfunctions $\{\psi_n\}$ of these operators with eigenvalues of physical quantities[2], i.e.,

$$\psi(x,t) = \sum_n C_n(t)\psi_n(x,t), \tag{12}$$

where

$$C_n(t) = \int \psi(x,t)\psi_n^*(x,t)dx \tag{13}$$
$$\text{и} \int \psi_n^*(x,t)\psi_m(x,t)dx = \delta_{n,m}.$$

Quantities $|C_n|^2$ represent the probability of state with the number $n$ in superposition (12).

In this case, the <u>principle of superposition in quantum mechanics</u> results from the correspondence of linear operators with different (orthogonality relations) eigenstates to physical quantities in an 'isotopic' space, with the mean values of these physical quantities $\langle f \rangle$ coinciding with the most probable $f_n$ eigenvalues (and exhibiting the highest probability of fluctuations).

The above-mentioned relationship between the space and the physical quantities determining the state of a system is observed in the gravitational theory. This role is played by the metric space, whose curvature is related to the energy-momentum tensor – the state of a physical system.

In the case of a quantum system, the space is an information Hilbert space of states with certain mean values of physical quantities that determine these states. The Schrödinger equation relates the physical quantities (energy, matter) to the information space in the same manner as the Einstein's gravitational equations relate the energy (matter) to the metric space.

Moreover, quantum states in the same way as in the gravitational theory (mass – curvature) are virtual, since their cumulative energy is equal to zero, as the wave functions correspond to E>0, and $\psi^*(x,t)$ – to E<0[3].

In other words, the states of a quantum system are merely possible states, with the probability of each of them being $|C_u|^2$. The meaning of $|\psi(x,t)|^2$ is the <u>maximum probability density of correlated fluctuations</u> (Δx-Δp), (Δt-ΔE), (Δθ-ΔM)…, generating the information waves $\psi(x,t)$, $\psi^*(x,t)$ <u>with the wave front coordinates – x, t</u>.

---

[2] Read on linear operators in any textbook on quantum mechanics
[3] Instead of time reversal

The space wherein all this occurs is triune: the Euclidean reference space (x, t), its reciprocal space of fluctuations ($k = \frac{2\pi}{\Delta x}$, $\omega = \frac{1}{2\pi \Delta t}$), and the information Hilbert ($\psi, \psi^*$), relating the latter two.

Motion characteristics of the particle itself: E, p, x, M,… are determined as mean values $\langle \psi | \hat{f} | \psi \rangle$ of the respective operators $\hat{f}$, which coincide with the most probable values $f_n$ – the spectrum of these operators.

The fluctuations proper are not localized $\Delta x \sim k^{-1}$, $\Delta t \sim \omega^{-1}$ and can possess variance $\omega(k)$, due to the correlation $\Delta x - \Delta t$. Under certain circumstances, e.g., during localization of a particle, $|\psi(x,t)|^2$ can represent its density (mass, charge,…), while its motion can be described by the Ehrenfest-type equations.

This motion of a quantum particle in the phase space can be represented as the motion of a pebble thrown into the water. Sinking, the pebble generates waves on the water surface, whose propagation carries the information on motion of the invisible pebble.

How can $\psi(x,t)$ describe real rather than virtual states? A possible scenario is similar to that of the black holes in the gravitational theory. In the same way as black holes absorb gravitation, they can absorb information associated with $\psi^*$, i.e., they contain a space of states for antiparticles. If black holes can be quantum anti-world carriers, then an increase in the entropy outside a black hole in the course of information absorption would compensate for the decrease in the entropy in the course of variation of the information associated with $\psi(x,t)$ and thus would make the quantum evolution reversible.

In this case, a relation between the gravitational (metric space), the physical state and the information quantum space for the regions longer than the Planckian length can be given by the following:

$$R = -\frac{8\pi k}{C^4} \langle \psi(x,t) | \hat{T} | \psi(x,t) \rangle -$$

an Einstein's equation, where R is the scalar curvature, in the brackets is the mean value of the energy-momentum operator for definite x, t corresponding to the metric space.

If $T^{(n)}$ are the eigenvalues $\hat{T}$, then $R^{(n)} = -\frac{8\pi k}{C^4} T^{(n)}$, and each of the n-states in the information space has a corresponding $R^{(n)}$ – the metric space curvature. In other words, in the metric space all possible states occupied by a particle to a probability of $|C_u|^2$ are determined by the 'quantum' curvature. Thus, all the three are combined here: metric space, matter and information (the Father, the Son, and the Holy Spirit).

The mean curvature would be $R = \sum_n |C_n|^2 R^{(n)}$ under the conditions of reversible evolution, and the space itself would consist of nested n-substates.

Another trajectory to realize the states is via the symmetry violation: a spontaneous violation by an external field, observation, or by an intrusion of the observer's mind (brain). The examples for the latter are: a mere omission of an equation for $\psi^*$ or various Gedanken experiments, such as the EPR-paradox, Bell's inequalities, etc.

### 3. Wavefunction phase.

Let us analyze the time dependence $\psi(t)$.

Write (9) as follows:

$$\frac{\partial \psi(t)}{\partial t} = -i\frac{E(t)}{\hbar}\psi(t) \tag{9`}$$

$$\text{и} \quad \psi(t) = \sum_k C_k(t)\psi_k(t). \tag{12`}$$

Substituting (12`) into (9`), we obtain

$$\sum_k \left\{ \dot{C}_k \langle n|k\rangle + C_k \langle n|\dot{k}\rangle + \frac{iE(t)}{\hbar}C_k \langle n|k\rangle \right\} = 0.$$

Form the solvability condition $\det|a_{nk}| = 0$, where $a_{nk} = \frac{\dot{C}_n}{C_k}\langle n|k\rangle + \langle n|\dot{k}\rangle + \frac{iE}{\hbar}\langle n|k\rangle$, the dot meaning $\frac{\partial}{\partial t}$, we obtain that if $|n\rangle$ and $|k\rangle$ are not orthogonal, then $C_n$ and $C_k$ are inseparable. This implies that the functions $\psi(x,t)$ cannot be presented by superposition, i.e., the unitary symmetry of the quantum system is violated. Therefore, <u>orthogonality</u> $\psi(x,t)$ is a necessary condition of discriminability of quantum states and unitary evolution of the system.

For the orthogonal states, from (9`) and (12`), multiplied scalarwise by $\langle n|$, we have

$$\dot{C}_k \delta_{nk} + C_k(\psi_n, \dot{\psi}_k) + \frac{iE_k}{\hbar}C_k \delta_{nk} = 0.$$

Solving this equation, we obtain

$$\begin{cases} n \neq k & 1 = e^{-\int_0^t (\psi_n, \dot{\psi}_k)dt'} & \text{a)} \\ n = k & C_n = e^{-\int_0^t \left\{(\psi_n, \dot{\psi}_n)dt' + \frac{iE_n(t')}{\hbar}\right\}dt'} & \text{b)} \end{cases} \tag{14}$$

In (14,b) $\int_0^t E(t')dt'$ is the dynamic phase of the wavefunction, for $E_n(t) = E_n$, being transformed into the steady-state phase. In the steady states $C_n = e^{\frac{iE_n t}{\hbar}}$ and $i\int_0^t (\psi_n, \dot{\psi}_n)dt' = 2\pi n$, where n are the integers. If $\psi$ depends on time implicitly $\psi(t) = \psi(q(t))$, then

$$i\int_0^q (\psi_n(q'), \frac{\partial}{\partial q}\psi_n(q'))dq' = \left|i\hbar(\psi_n(q'), \frac{\partial}{\partial q}\psi_n(q') = p_n\right| =$$

$$= \int_0^q p_n(q')dq' = 2\pi\hbar n$$

is the Bohr-Sommerfeld quantization rule.

Quantity $\int_0^t (\psi_n, \dot{\psi}_n)dt'$ is the geometrical (topological) phase of the wavefunction. Note that $(\psi_n, \dot{\psi}_n)$ is imaginary and $(\psi_n, \dot{\psi}_n) = -(\dot{\psi}_n, \psi_n)$; for the real $\psi_n$, $(\psi_n, \dot{\psi}_n) = 0$. Denoting $(\psi_n, \dot{\psi}_n) \equiv A(E)$, we obtain $A(E) = A^*(-E)$. From (14,a) follows that

$$i\int_0^t (\psi_n(t'), \psi_k(t'))dt' = 2\pi m\hbar, \qquad (15)$$

where m=0,1,2,… are integers.

If quantum numbers remain invariant in time (the Ehrenfest adiabatic invariant), then differentiating (15) with respect to t, we have

$$(\psi_n(t), \dot{\psi}_n(t)) = 0 \qquad (16)$$

– the condition of wavefunction fixation (Born-Fock condition) under adiabatic conditions –
$$\left|\frac{(\psi_m|\dot{H}|\psi_n)}{E_n - E_m}\right| \ll 1.$$

Let us pick out the wavefunction phase in an explicit manner, representing
$$\psi_n(t) = \psi_n(q(t)) = e^{i\theta(q(t))}\varphi_n(q(t)) \qquad (17)$$

Let the time $t$ range in a certain interval $0 \leq t \leq t_0$. Then, for $(\varphi_n, \varphi_k) = \delta_{nk}$ from (17), (16), we obtain

$$\dot{\theta}^{(n)}(q) = i\left(\varphi_n, \frac{\partial \varphi_n}{\partial q}\right)\dot{q}. \qquad (18)$$

Denote
$$i\left(\varphi_n, \frac{\partial \varphi_n}{\partial q}\right) \equiv A^n, \qquad (19)$$

then
$$\dot{\theta}^{(n)} = A^n \dot{q} \qquad (20)$$

and
$$A^n = \frac{\partial \theta^{(n)}(q)}{\partial q}. \tag{21}$$

If the wavefunction phase is not constant, its derivative would determine the gauge field $A^n$. In the case of an electrically charged particle, $A^n$ is the electromagnetic-field vector potential that transforms as $\dot{A}^n = A^n + \frac{\partial \alpha_n}{\partial q}$, where $\frac{\partial \alpha_n}{\partial q}$ is the potential gauge.

In terms of (18) – (21), the phase proper is given by

$$\theta^{(n)}(t_0) = \int_0^{t_0} \dot{\theta}^{(n)} dt = \int_C A^n dq = \theta^{(n)}(q(t_0)) - \theta^{(n)}(q(0)). \tag{22}$$

If the motion is cyclic and the loop C is closed, then

$$\theta^{(n)}(t_0) = \oint_C A^n dq \tag{23}$$

and the phase is determined by the closed loop only, implying a topological Berry's phase [23]. Using the Stokes theorem, we obtain

$$\theta^{(n)} = \int_S \vec{F} d\vec{s}, \tag{24}$$

where ds is the surface element stretched over the loop C, and $F^{(n)} = rot\vec{A}^{(n)}$ is the gauge field strength.

a) If we consider an electromagnetic field, then it is the circulation $A^n$ and (24) which determine the magnetic flux $\Phi^{(n)}$, and the phase has the meaning of a magnetic flux that, according to (15), is quantized – $\Phi_0 = (2\pi m\hbar)$.

б) If $\varphi_n$ -are the wavefunctions of elastic quantum atomic displacements in the crystals, then $A^n$ would be the gauge field of linear defects – total displacement, and $\theta^{(n)}$ would have the meaning of the Burgers vector ~$na$ ($a$ – lattice constant) and $F_{ik}^{(n)} = \frac{\partial A_k^{(n)}}{\partial q_i} - \frac{\partial A_i^{(n)}}{\partial q_k}$ would be the dislocation density tensor. In both cases, $\theta^{(n)}$ will be a <u>quantized macroscopic observable quantity.</u>

From the formulas for $F^{(n)}$ and $A^{(n)}$, we have

$$F^{(n)} = \nabla x A^{(n)} = \text{Im}(\nabla \varphi_n \times \nabla \varphi_n).$$

Using for $\nabla \varphi_n$ the first-order perturbation theory

$$\nabla \varphi_n = \sum_{m \neq n} \frac{(\varphi_m | \nabla A | \varphi_n)}{E_m - E_n} + (\varphi_n | \nabla \varphi_n | \varphi_n),$$

we obtain
$$F^{(n)} = \text{Im} \sum_{m \neq n} \frac{(n|\nabla H|m) \times (m|\nabla H|n)}{(E_m - E_n)^2}.$$

(25)

Quantity $F^{(n)}$ is singular for $E_m = E_n$ – degeneracy of the spectrum, and is a geometric (topological) analog of the magnetic field strength (H – Hamiltonian of the system).

## 4. Macroscopic quantum effects.

Calculate the quantum electrical current using a well-known formula

$$j = \frac{i\hbar e}{2m}(\psi \nabla \psi^* - \psi^* \nabla \psi) \qquad (26)$$

Using (17) and (21), we obtain

$$j = \frac{\hbar}{2m} 2e \nabla \theta^{(n)} + \frac{i\hbar e}{2m}(\varphi_n \nabla \varphi_n^* - \varphi_n^* \nabla \varphi_n) \qquad (27)$$

Let us consider the first term in (27). Since $\nabla \theta^{(n)}$ from (21) represents $\vec{A}$, then, denoting this term as $j_s$, we arrive at

$$\vec{j}_s = 2e \frac{\hbar}{2m} \vec{A}. \qquad (28)$$

The contribution $j_s$ to the total current is determined by the electromagnetic-field vector potential, with the charge transferred being $2e$. The current $j_s$ exists for E=0 (A≠0). However, from the Ohm's law $\rho \vec{j} = \vec{E}$ follows that for E=0, j≠0 only for ρ=0 (σ→∞). If under these conditions the magnetic field does penetrate into the conductor, then it is an ideal metal, if not so (Meissner effect) then the specimen is a superconductor. For this current, in addition to (28) from the condition $\vec{E} = \frac{1}{C}\frac{\partial \vec{A}}{\partial t}$ we have

$$\frac{\partial \vec{j}}{\partial t} \sim \vec{E}, \qquad (29)$$

which implies that the steady-state current is not related to $\vec{E}$, both (28) and (29) are the London equations for superconductors. It should be emphasized that similarly to subsections a) and b) of the previous section, the formulas (28) and (29) describe the <u>real macroscopic quantum current</u>, while (26) includes a wavefunction that bears the meaning of information.

Superconducting current is carried by pairs of electrons. These pairs are condensed into a macroscopic superconducting state with violated gauge symmetry. This state is a new <u>thermodynamic phase</u> of a metallic material characterized by a quantum magnetic flux $\sim n\Phi_0$. Note that the Meissner effect corresponds to n=0, where the wavefunction determining $\vec{A}$ cannot discontinue, and the trajectory of integration $\int A^{(n)} dq$ of the wavefunction phase (Aharonov-Bohm) can collapse into a point. While the phenomenon of superconductivity has been studied in detail, the issue of <u>interdependence</u> between the wavefunction in quantum mechanics and that in macroscopic description is still open. In view of this circumstance, the wavefunctions in the Hilbert space fail to describe the superconducting state, since they already carry the maximum information on the initial quantum system corresponding to the lowest entropy. A further decrease in the entropy $\Delta S = S_s - S_n < 0$ in the course of transition into a superconducting macroscopic state is only possible only, provided there is certain <u>interdependence</u> between the quantum information space and the macroscopic physical space. This interrelationship (a certain decoherence of the initial state) reduces symmetry of the Hamiltonian, making it non-invariant with respect to the gauge transformations (rotations in the 'isotopic space'), while leaving invariant the Ginzburg-Landau thermodynamic potential in the transitional current – $T_c$.

Note that $\Delta S=0$ in T=0 and $T=T_c$. This interdependence results in localization of two electrons within their 2x bond length, so that for every particle density

$$\hat{\rho}\psi(x') = e\delta(x-x')\psi(x')$$
$$\langle\rho\rangle = \langle x|\hat{\rho}|x\rangle = e|\psi(x)|^2$$

is the real mean charge density in point x – n(x).

This 'localization' – decoherence – suppresses fluctuations $\Delta x$, so that the fluctuations of the charge density proper $\Delta n(x) \gg \Delta x$. Further on, the Bose-Einstein condensation results in macroscopic $\psi(x)$ and a macroscopic charge – 2Ne, forming a new long-range order – a superconducting phase. The superconducting (superfluid) current associated with the phase condensate inhomogeneity is a macroscopic quantity.

The formula for current nominally coincides with (26), however, $\psi$ is a macroscopic order parameter, and it is its dynamics (Landau – Khalatnikov equations) which determines the phase transition into a new <u>thermodynamic state</u>. All these considerations refer to paired electrons only. The rest of the electrons would remain in the former microscopic state and, hence, the fewer microparticles, the more cooperative the macroquantum state will be.

Thus, pair production (decoherence of microstates) and condensation of the paired fraction of the electrons changes the <u>roadmap</u> of a possible behavior of microparticles into a real macroscopic state.

This state has a corresponding value of n=0, and the <u>roadmap</u> has the corresponding n=1,2,.... T the temperatures lower than $T_c$ (close to zero), presence of the states with n=1,2,… and, accordingly, the magnetic field $\vec{B}$ (Aharonov – Bohm phase) destroys the superconducting state by reducing the number of states with n=0. "Localization" of a quantum object makes it observable, and the fluctuations $\Delta x \to 0$ and wavefunctions in the information space become discontinuous. It is these discontinuities: for the density $\delta\psi(x) = \psi_+ - \psi_-$ and displacements $\delta U = U_+ - U_-$, which determine the superfluid motion. The macroscopic quantum objects have their respective discontinuities in the Hilbert information space. Within these discontinuities, $\psi(x)$ are undefined and $\delta\psi(x)$ are defined, so that $\dfrac{\partial^2 \psi(x)}{\partial x_i \partial x_j} \neq \dfrac{\partial^2 \psi(x)}{\partial x_j \partial x_i}$ and particularly $\delta\psi$ prescribe the system's behavior either as a phase transition resulting from high-temperature fluctuations of the order parameter, or as Abrikosov-Shubnikov vortices, or as dislocations in the case of displacements U(x), or as a quantum phase transition [25].

It is critical that this happens during suppression of fluctuations $\Delta x$. Getting a little ahead (of the discussion of interference), let us point out that in essence $\Delta x$ is a possible 'trajectory difference' during the probability wave interference. For coherent states, from the maximum interference condition $\Delta x = n\lambda$ it is clear that for $n \to \infty$ $\lambda \to 0$ is a classical limit. However n=0 is only possible for $\Delta x = 0$, i.e., without any fluctuation of coordinates. In this state, $\lambda$ can be nonzero, and this wavelength corresponds to a cooperative macroscopic quantum state. For $\vec{B}=0, \rho=0$ is the superconducting state. Also note that n=1 corresponds to the maximum interference $|\psi|^2$.

### III. Quantum interference and corpuscular-wave dualism.

Let us briefly summarize the above before proceeding to discuss interference.

1. Quantum mechanics is a description of space-time fluctuations determining the motion of a quantum object in a homogenous physical ST. Due to conservation of their homogeneity, the fluctuations cannot be localized and posses wave nature: fluctuations of coordinates represent a wavelength, and fluctuations of time represent its frequency, forming the 'reciprocal space'. As any other, these fluctuations are determined by probability and entropy; hence it is natural that the waves representing these fluctuations are the waves of probability – information waves.

2. Information waves can be represented by wavefunctions in the 'isotopic' Hilbert information space.

3. Production of entropy, associated with the probability of fluctuations, under adiabatic conditions and in the presence of an adiabatic invariant prescribes the functionals in this space, which correspond to physical quantities in an ordinary ST. These functional are by definition linear Hermitian operators, and their mean values represent the respective physical quantities.

4. To determine the wavefunctions, a (Schrödinger) equation is constructed, which combines the determination of the energy operator in the space of wavefu7nctions and the explicit form of the operator represented via the physical quantity proper, according to the correspondence principle.

5. The wavefunctions represent a 'roadmap' (navigator) for a possible motion of a quantum particle. Naturally, the interfering wave (waves) also describes possible motion. What causes the quantum particle to follow the 'roadmap'? The answer is as follows.

6. The above relation (14,b) derived for coefficients $C_n(t)$ during adiabatic evolution determines them via the topological Berry's phase. In a more general case ($\dot{q} \neq 0$), the equations for $C_n(t)$ are given by

$$\dot{C}_n(t) = i \sum_{m \neq n} C_m \gamma_{nm} e^{-i \int_0^t (E_m - E_n) dt'} \tag{30}$$

and the respective equations for $C^*_n(t)$.

Here $\gamma_{nm} = -i\int \psi^*_n \dfrac{d}{dt}\psi_m dq$ is the Hermitian matrix. Equations for C and C* represent a Hamiltonian system of equations in a classical sense, with the Hamiltonian function

$$H(C,C^*,t) = -i\sum_{n,m}\gamma_{nm} C^*_n C_m e^{-i\int_0^t (E_m - E_n)dt'}, \qquad (31)$$

describing a classical distributed system [26], which can be readily shown when going from variables $C_n, C^*_n$ to the variables: action - $I_n$ and angle - $\alpha_n$: $C_n = \sqrt{I_n} e^{-i\alpha_n}$; $C^*_n = \sqrt{I_n} e^{i\alpha_n}$; $I_n = |C_n|^2$, $\alpha_n = -arctg\dfrac{C_n - C^*_n}{i(C_n + C^*_n)}$ [26,27].

Instead of (30) and (31), we have

$$H(H,\alpha,t) = \sum_{n,m}\gamma_{n,m}\sqrt{I_n I_m} e^{[\alpha_m - \alpha_n + (E_m - E_n)t]} \qquad (32)$$

and

$$\dot{I}_n = 2\sum_{m=0}\sqrt{I_n I_m}\gamma_{nm} e^{-i[\alpha_n - \alpha_m + (E_m - E_n)t]}$$

$$\dot{\alpha}_n = 2\sum_{m=0}\sqrt{\dfrac{I_n}{I_m}} \mathrm{Re}\left\{\gamma_{nm} e^{-i[\alpha_n - \alpha_m + (E_m - E_n)t]}\right\} \qquad (33)$$

The above system of equations (33) is a classical Hamiltonian system [26,27]. This implies that coefficients C, C* prescribe the 'trajectory' of a quantum particle in the information space, along which 'classical' motion takes place in the same way as a classical particle moves along its phase trajectory (19). Note that it is $|C_u|^2 = I_n$ which corresponds to action, and its variation $|C_u|^2$ in time corresponds to undetermined (nonfixed) energy during this motion that can be represented by a superposition of states.

Let us address quantum interference using two slits (Fig. 1).

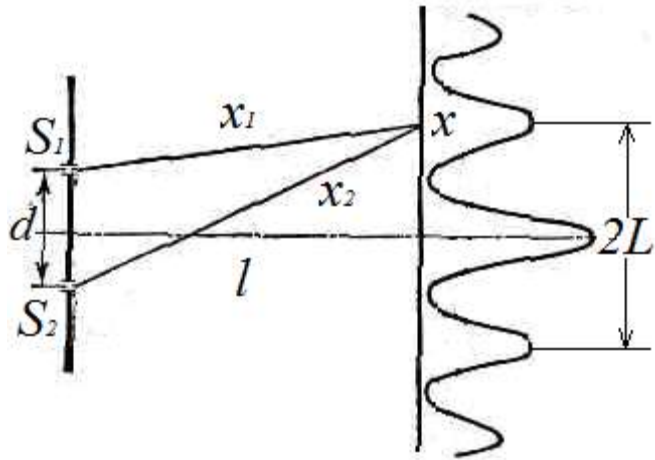

Fig.1. Interference of an information wave in two slits

The screen length – 2L, spacing to the screen – $l$, spacing between the slits – $d$, particle position on the screen – x.

Given two slits, $S_1$, $S_2$, the wavefunction is given by $\psi = \psi_1 + \psi_2$, where

$$\psi_1 = a_1 e^{2\pi i(\frac{x_1}{\lambda} - \frac{t}{T})}; \quad \psi_2 = a_2 e^{2\pi i(\frac{x_2}{\lambda} - \frac{t}{T})}.$$

The term $|\psi|^2$ includes an interference term, which increases (or decreases) the intensity $\sim \cos\frac{2\pi(x_1 - x_2)}{\lambda}$, where

$$x_1 - x_2 = \Delta x \simeq \frac{xd}{l} \text{ is the trajectory difference.}$$

Given the quantum motion of a particle, the trajectory difference corresponds to uncertainty (fluctuations) of the coordinates $\Delta x \geq \frac{h}{p}$. For coherent states, it is these fluctuations which determine a possibility of trajectory difference even for a single particle. Condition $\Delta x = n\lambda$ determines the appearance of interference maxima and the relationship between the information wavelength $\lambda$ and momentum $\lambda_{max} = \frac{h}{p}$.

Thus, uncertainty $\Delta x$ represents a possible trajectory difference and, respectively, the fringe pattern of the information waves. The particle position on the screen cannot be determined from the presented 'roadmap' as $\langle x \rangle = \langle \psi | x | \psi \rangle$, since for an exactly determined momentum $\langle x \rangle = 0$. This position is determined by the relation $\Delta x = \frac{xd}{l}$, for $\Delta x = n\lambda$ and $\lambda_{max} = \frac{h}{p}$, $p = \sqrt{2mE}$, $E \to eU, kT, ...$

In other words, $\langle x \rangle$ is prescribed by the momentum (energy) of a prepared state of the particle participating in the experiment and by the experimental parameters.

Any circumstance, resulting in a certainty such that $\Delta x \to 0$, would cause the interference to disappear. Indeed, for $\Delta x \to 0$ x is defined and $\Delta p \neq 0$ (which violates space homogeneity). This gives rise to $\Delta \lambda \sim h/\Delta p$ and results in decoherence of the information waves – extinction of the fringe pattern. In the general case, $x = \frac{h - \Delta x p}{\Delta p}$ and under the condition of interference, where the trajectory difference $\Delta x = \frac{xd}{l}$, we have

$$x = \frac{h}{\Delta p + \frac{d}{l}p} = \frac{h}{\Delta P}. \tag{34}$$

An observation with a device (measurement) introduces an additional uncertainty $\Delta p' = \frac{d}{l} p$ into the momentum. In other words, without an intrusion of 'measurement', localization can only take place for $\Delta x \to 0$, while the measurement introduces an uncertainty $\Delta p'$ into the momentum, and even for $\Delta p = 0$ results in localization of a particle with $x = \frac{h}{\frac{d}{l} p}$; the particle position is now determined by the measurement (instrument) parameters and the initial prepared state (p) of the quantum particle.

It is appropriate here to make a few comments:

1. Recurring to the binding entropy $\Delta S$ responsible for quantum correlations, we see that it is impossible to rile them out within the quantum framework: $\Delta S \neq 0$ under no circumstances. In other words, it is impossible to localize a particle without an external factor (violation of symmetry, measurement, "offside" interaction) and quantum description is, therefore, <u>nonlocal</u> in its nature: either quantum correlations or object localization. It should be underlined that localization takes place in a physical space. This means that any measurement fixes the particle in a physical rather than "isotopic" space. In fact this is the wavefunction (information) collapse in the course of measurement transferring <u>information into knowledge</u>. Localization causes a transition of information into a "real" observable physical characteristic of an object. This transition is actually a transfer from a representation via linear operators and <u>superposition of information waves to a given physical quantity</u> – its mean value, and in the case of eigenstates of these operators – to specifying the physical state itself. Clearly, the information does not directly turn into matter, but unambiguously points to where the density of the latter is concentrated.

This implies that it is not matter which manifests wave properties in the macroworld, but the characteristics of its motion (particles) – fluctuations of the space-time coordinates. These fluctuations (as well as any other) are described by their probability that, due to the wave behavior of fluctuations themselves, can be represented by probability waves (wave process) in the "isotopic" information space – time. The wave characteristics of fluctuations, $\Delta x, \Delta t$ - wavelength and frequency, in their turn, are determined by the momentum and energy of "prepared" states taking part in the particle motion.

For $\Delta x$ the relation is $\Delta x \geq \frac{h}{p}$ and $\Delta x = \lambda$ (as well as for $\Delta t$). This allows us to argue that the equalities $\frac{h}{p} = \lambda$, $h\nu = E$ are due to the wave character of fluctuations, and in no way to the relations validating the wave properties of matter. Matter has no wave properties. In other words, the corpuscular-wave dualism consists in the <u>duality</u> of fluctuations of space-time coordinates: on the one hand (for coherent states) $\Delta x = h/p$, due to the fact that $h = \min \Delta s$ (action) and the physical space is homogenous, and, on the other hand, the relation $\Delta x = \lambda$ determines the trajectory difference even for a single particle (!) and the respective condition for the probability wave interference – information. This wave behavior is objectively nonlocal due to quantum correlations, and represents a virtual state of an object – information about it, and the measurement (external environment) only can localize this object, and add "real features" to it and the knowledge about them.

From our standpoint, we should differentiate between the information about an object and the knowledge about it. Information is an objective measure of cognizability of an object or a phenomenon (a possibility of obtaining knowledge).

Knowledge is a subjective measure of cognizance of an object or phenomenon by an observer.

The process of measurement lies between the two. This process also involves an additional change of the entropy $\Delta S_{изм}$, which, by changing the binding entropy $\Delta S$, destroys the quantum correlations and localizes the particle, possibly transferring it into a conventional object. A measure of this transfer would be the residual entropy [20] equal to $\Delta S - \Delta S_{mesur.}$ It is the nulling of $\Delta S - \Delta S_{mesur.}$, which determines this transformation.

Apart from a physical phenomenon (system) there is information about it and a possibility for an observer to record this information – to obtain knowledge about this phenomenon.

An example showing the difference of information is the following.

Let a quantum particle have the angular momentum $M^2 = M_x^2 + M_y^2 + M_z^2 = l(l+1)$, $(\hbar = 1)$. For l=1, M=2 the three projections of the momentum have a corresponding magnetic quantum number $m = 1, 0, \bar{1} \to x, z, y$.

Now, let us measure M. In accordance with the projection postulate, this measurement has a corresponding projector $P_m = |m \times m|$. We want to determine $M_z$. Prior to the measurement, however, all the three coordinate axes are equivalent. There is no doubt that information about $M_z$ does exist, but in order to obtain it (turn it into knowledge), one has to fix the system of coordinates. In other words, the system of coordinates (reference system) has to be fixed for the observer to obtain knowledge.